\documentclass[prb,  twocolumn, showpacs, superscriptaddress,longbibliography]{revtex4-1}
\usepackage{amsmath, amsfonts, amssymb, mathrsfs, mathtools, stmaryrd}
\usepackage{graphicx, color}
\usepackage{xargs}                      
\usepackage[pdftex,dvipsnames]{xcolor}  
\usepackage[caption=false]{subfig}
\usepackage{tikz}
\usepackage{float}
\usepackage{tabularx}
\usepackage{booktabs}
\usepackage{bm}
\usepackage{braket}
\usepackage{epstopdf}
\usepackage{dsfont}
\usepackage{float}
\usepackage{mathtools}
\usepackage{hyperref}
\hypersetup{
   colorlinks,
   citecolor=blue,
   filecolor=blue,
   linkcolor=blue,
   urlcolor=blue,
   breaklinks=true
}
\graphicspath{{Figures/}} 

\definecolor{grey}{rgb}{.6,.6,.6}

\begin{document}
\title{Numerical investigation of the structure factors of the Read-Rezayi series}
\author{Lo\"ic Herviou}
\affiliation{Univ. Grenoble Alpes, CNRS, LPMMC, 38000 Grenoble, France}
\affiliation{Institute of Physics, Ecole Polytechnique F\'ed\'erale de Lausanne (EPFL), CH-105 Lausanne, Switzerland.}
\author{Frédéric Mila}
\affiliation{Institute of Physics, Ecole Polytechnique F\'ed\'erale de Lausanne (EPFL), CH-105 Lausanne, Switzerland.}
\begin{abstract}
We numerically investigate the guiding center stucture factors of several states in the Read-Rezayi family. Using exact diagonalizations on the torus and density matrix renormalization group on an infinite cylinder, we test a conjecture proposed in [\href{http://dx.doi.org/10.1103/PhysRevLett.113.046803}{Can et al, Phys. Rev. Lett. 113 (2014)}] for the $\nu = \frac{3}{5}$ and $\nu = \frac{4}{6}$ Read-Rezayi states. Furthermore, we discuss the strong finite-size effects present in numerically accessible wavefunctions, and provide a simple recipe to minimize them on manifolds where non-Abelian theories have topological degeneracies.
\end{abstract}
\maketitle
\section{Introduction}

The fractional quantum Hall (FQH) states are topological phases of matter characterized by the emergence of exotic quasiparticles with unusual charge and exchange statistics\citep{Moore1991}.
In principle, their topological properties can be probed through the determination of the Hall conductance\citep{Tsui1982, Laughlin1983}, the Hall viscosity\citep{Avron1995, Haldane2009, Read2009, Read2011} --- connected to the so-called topological spin\citep{Zaletel2013, Tu2013} --- and the chiral central charge\citep{Kane1997} of the edge theory.
While the Hall conductance itself is a simple function of the density, the latter two quantities provide a more complete characterization of the topological order\citep{Wen1990, Kitaev2003}.
They prove nonetheless difficult to estimate in the numerical wavefunctions one has access to, due to the restricted system sizes and the edge effects.\\

Structure factors (SF) are a standard way to characterize the properties of gapped and gapless phases of matter.
In FQH systems, instead of considering the full motion of the electrons, it is possible to separate the orbital motion induced by the geometry of the Landau levels from the motion of their center of mass, the guiding centers\citep{Haldane2011, Park2014}.
The SF of the guiding centers\citep{Girvin1986, Haldane2009} carry information on the fractional topological order.
Firstly, the bulk gap of topologically ordered states implies that the SF should decay as the fourth power of the momentum at long wavelength\citep{Girvin1986} so that their expansion takes the form
\begin{equation}
\overline{S}(\vec{k}) = \overline{S}_4 l_B^4 k^4 + \overline{S}_6 l_B^6 k^6 + ...\label{eq:formSF}
\end{equation}
Secondly, the coefficient $\overline{S}_4$ of the dominant term should satisfy a lower bound connected to the topological spin and the Hall viscosity\citep{Haldane1983, Golkar2016, Nguyen2021, Nguyen2022}.
It has been conjectured --- and analytically and numerically verified for the Laughlin\citep{Laughlin1983} and Moore-Read\citep{Moore1991, Read1992} states --- that this bound is saturated for maximaly chiral conformal wavefunctions\citep{nguyen2014lowest, Kalinay2000, Wang2019, Dwivedi2019, Kumar2023}.
Finally, for this class of functions, the next-order coefficient $\overline{S}_6$ has been conjectured to be a simple function of the chiral central charge of the edge theory\cite{Can2014, Can2015, Gromov2017, Gromov2017b}.\\

In this paper we investigate these structure factors in the Read-Rezayi\citep{Read1999} family of states, with two main objectives.
Firstly, we verify the relations derived for $\overline{S}_4$ and $\overline{S}_6$ on more complex non-Abelian conformal states, the third and fourth states in the Read-Rezayi family.
Secondly, we investigate the finite-size effects present in numerically achievable systems, i.e. wavefunctions obtained from actually solving a Hamiltonian problem, and the effect of topological degeneracies on the guiding centers structure factors.
The analytical computations in Refs.~\onlinecite{Can2014, Can2015, Gromov2017, Gromov2017b} were formally performed in the bulk of an infinite system.
Then, the different non-Abelian FQH groundstates are locally indistinguishable and the guiding center SF are neither manifold nor state dependent.
Conversely, numerical computations can be done on a finite torus or a semi-infinite cylinder where the groundstate is not unique and the states are not indistinguishable.
We propose a simple recipe to minimize size-effects and recover the theoretical results at cheaper numerical costs.
Despite being limited to relatively small systems, our formula is enough to extract accurate estimates of $\overline{S}_4$ and $\overline{S}_6$ in agreement with the theoretical predictions.
Our numerical results are summarized in Table~\ref{tab:PropertiesRR} and Fig.~\ref{fig:AllGCSF}.\\

The paper is organized as follows. Sec.~\ref{sec:definitions} gives the global definitions of the guiding centers and their structure factors.
Sec.~\ref{sec:NumMet} covers the numerical methods used to extract the states of the Read-Rezayi series from their parent Hamiltonians and compute their properties. 
We also detail how to obtain more accurate estimates of the structure factors for non-Abelian theories.
Finally, Sec.~\ref{sec:SFRR} studies the properties of the structure factors in the Read-Rezayi series.


\section{Structure factors of the guiding centers}\label{sec:definitions}
\subsection{Electronic guiding centers}
The guiding centers\citep{Girvin1986, Haldane2009} are a convenient way to describe the trajectories of electrons in a magnetic field.
The orbital motion of the electron, which depends on the Landau level (LL) in which it resides, is separated from the global trajectory of its center of mass --- the guiding center.
Mathematically, the guiding center position operator $\vec{R}$ is defined as
\begin{equation}
R^a  = r^a + \varepsilon^{a, b} \pi_b,\qquad [R^a, R^b] = - i l_B^2 \varepsilon^{a, b}, 
\end{equation}
where $\pi_a = p_a + \frac{e}{c}A_a$ are the magnetic momenta and $l_B = \frac{\hbar}{eB}$ the magnetic length.\\

Already, this separation leads to a stark contrast between integer and fractional quantum Hall states.
The guiding centers of the integer quantum Hall states organize trivially, as the LLs are either fully occupied or empty.
There, topology simply arises from the geometry of the LLs and $\overline{S}_4$ and $\overline{S}_6$ trivially vanish.
Conversely, the exotic topological properties of the FQH states are necessarily connected to the behavior of the guiding centers.
The structure factors of the guiding centers are therefore a relevant probe of topological order.

\subsection{Definition and properties of the structure factors}
In the rest of the paper, for the sake of simplicity, we focus on the case of a single (partially occupied), spin-polarized LL.
We denote $N_e$ the number of electrons within the LL of choice, and $N_\phi$ the number of orbitals forming the LL, i.e., the number of magnetic fluxes threading the system.
The filling is denoted as $\nu$.
Formally, the following formulas are valid on a periodic system (a torus), but can be adapted to the infinite cylinder.
The guiding center structure factors are simply defined as
\begin{equation}
\overline{S}( \vec{q}) = \frac{1}{2 N_\phi} \langle \{ \delta\overline{\rho}(\vec{q}), \delta\overline{\rho}(-\vec{q})\} \rangle \label{def_SF}
\end{equation}
where
\begin{equation}
\overline{\rho}(\vec{q}) = \sum\limits_{j = 1}^{N_e} e^{i \vec{q}. \vec{R}_j} \text{ and }
\delta\overline{\rho}(\vec{q}) = \overline{\rho}(\vec{q}) - \langle \overline{\rho}(\vec{q}) \rangle.
\end{equation}
The guiding center density $\overline{\rho}(\vec{q})$ obeys  the Girvin-Macdonald-Platzman\citep{Girvin1986} algebra
\begin{equation}
[\overline{\rho}(\vec{k}), \overline{\rho}(\vec{q})] = 2 i  \sin\left(\frac{1}{2}\varepsilon^{a, b} k_a q_b  \right) \overline{\rho}(\vec{k} + \vec{q}).
\end{equation}

With these definitions, the guiding center structure factors satisfy a few remarkable relations.
First, they are directly connected to the conventional structure factors by the convolution\citep{Haldane2009}
\begin{equation}
\nu (S(\vec{q}) - 1) = \vert f_l(\vec{q}) \vert^2 ( \overline{S}(\vec{q}) -  \overline{S}(\infty)), \label{eq:RelSs} \end{equation}
where
\begin{equation}
S( \vec{q}) = \frac{1}{2 N_e} \langle \{ \delta\rho(\vec{q}), \delta\rho(-\vec{q})\} \rangle, \quad
\rho(\vec{q}) = \sum\limits_{j = 1}^{N_e} e^{i \vec{q}. \vec{r}_j},
\end{equation}
\begin{equation}
f_l(\vec{q})  =  \vert \langle m, l \vert e^{i \vec{q}.\vec{r}} \vert m, l \rangle \vert,
\end{equation}
\begin{equation}
\langle m, l \vert e^{i \vec{q} . \vec{r}} \vert n, l \rangle = \langle m, 0 \vert e^{i \vec{q} . \vec{R}} \vert n, 0 \rangle  f_l(\vec{q})\label{eq:CompGC}.
\end{equation}
Here, $\vert m, l \rangle$ is a fermionic orbital in the $l^\mathrm{th}$ LL under study.
The form factor $f_l$ generally decays as a Gaussian at long distance.
It captures the geometrical properties of the  $l^\mathrm{th}$ Landau level.
The existence of the infinite limit $\overline{S}(\infty)$ technically implies a rotation invariance at short distances.
For fermions, it should be equal to
\begin{equation}
\overline{S}(\infty) = \nu (1 - \nu).
\end{equation}
The convolution in Eq.~\eqref{eq:RelSs} shows that the same information is in principle present in the conventional and guiding center structure factors.
However, extracting $\overline{S}_4$ and $\overline{S}_6$ from $S$ would require careful substractions of dominant irrelevant terms which quickly spoil the numerical precision.\\

Secondly, the key property of the guiding center structure factors\citep{Girvin1986}  is that they verify
\begin{equation}
\overline{S}(x \vec{q}) \propto x^4
\end{equation}
for incompressible FQH groundstates with inversion and translation symmetry.
The expression relaxes to Eq.~\eqref{eq:formSF} if rotation invariance is assumed.
%

\subsection{Structure factors of conformal states}
FQH states can be characterized through several ``topological'' quantities.
In the composite-boson picture promoted by Jain\citep{Haldane1983, Halperin1984, Jain1989, BookHalperinJain}, the elementary particles are made of a set of $\tilde{p}$ electrons occupying $\tilde{q}$ orbitals, at a filling $\nu = \frac{\tilde{p}}{\tilde{q}}$.
Typical examples are the Laughlin states\citep{Laughlin1983} at filling $1/m$ with $\tilde{p} = 1$ and $\tilde{q} = m$, or the fermionic Moore-Read\citep{Moore1991, Read1992} state with $\tilde{p}=2$ and $\tilde{q}=4$.
These quantities are not enough to fully characterize a FQH states.\\

Of interest for the current paper are the topological spin and the edge central charge.
The topological spin $s$ is related to the non-trivial braiding statistics of the quasiparticles.
It measures the difference in angular momentum of the effective composite boson and a uniform grouping of $p$ electrons in $q$ orbitals.
It is related to the "shift" $\mathcal{S}$ of the FQH states.
Take the uniform FQH state on a sphere enclosing a magnetic monopole of charge $N_\phi$.
The shift is defined as
\begin{equation}
\mathcal{S} = \nu^{-1} N_e - N_\phi.
\end{equation}
It verifies
\begin{equation}
\mathcal{S} = 2\overline{s}, \text{ with } \overline{s} = (l + \frac{1}{2})  - \frac{s}{p},
\end{equation}
where $l$ is the Landau level under study and $\nu = \frac{p}{q}$ with $p, q$ coprime.
It is also related to the Hall viscosity\citep{Avron1995, Haldane2009, Read2009, Read2011} through
\begin{equation}
\eta_H = \frac{\hbar \nu}{4 \pi l_B^2} \overline{s}.
\end{equation}
The Hall viscosity characterizes the response due to the correlations of a FQH fluid to shear deformations.
Quite naturally, it will appear as a bound in the structure factor\citep{Haldane1983, Golkar2016, Nguyen2021, Nguyen2022}:
\begin{equation}
\overline{S}_4 \geq \frac{\pi l_B^2}{\hbar} \vert \eta_H \vert = \frac{\nu \vert s\vert}{4}.\label{eq:defS4}
\end{equation}
Intuitively, one can understand this bound as the contribution to the structure factors coming from the Berry curvature of the quantum state.
This bound is expected to be saturated for maximally chiral states\citep{nguyen2014lowest, Kalinay2000, Wang2019, Dwivedi2019, Kumar2023}.\\

Edge modes of ideal FQH states are described by $1+1$ conformal field theories, and therefore can be characterized by their central charge.
Here more precisely, the quantity of interest is the signed central charge
\begin{equation}
\tilde{c} = \sum\limits_{e\text{ edge modes}} \varepsilon_e c_e,
\end{equation}
where $\varepsilon_e = 1$ (resp. $-1$) if the edge-mode propagates (resp. counterpropagates).
It has been conjectured that\cite{Can2014, Can2015, Gromov2017, Gromov2017b}, for exact conformal states, 
\begin{equation}
\overline{S}_6 = \frac{\nu \vert s\vert}{8}\left( \vert s\vert - \frac{\tilde{c} - \nu}{12\nu \vert s\vert}\right). \label{eq:defS6}
\end{equation}
This conjecture has been verified analytically or numerically for the Laughlin and Moore-Read states on the sphere or the cylinder\citep{nguyen2014lowest, Kalinay2000, Wang2019, Dwivedi2019}. 

\section{Numerical methods and structure factor analysis}\label{sec:NumMet}
In this section, we briefly describe our numerical methods, and the associated technical challenges.

\subsection{Exact diagonalization on the torus}\label{sec:ED}
We study FQH states on a torus using exact diagonalization.
The torus allows us to work with translation-invariant wavefunctions, at the price of some technical and conceptual difficulties.
For example, the boundary conditions of the torus implicitly break the rotation symmetry at the long distances we are interested in.
It is in principle possible to study the guiding center SF on a sphere.
However, the number of accessible momenta is drastically reduced so that the extraction of $\overline{S}_4$ and $\overline{S}_6$ is challenging for the more complex states considered in this paper.\\

More precisely, we consider a (twisted) torus of aspect ratio $r = L_x / L_y$ and denote by $\tau$ the twist parameter, such that the system is invariant by translation under $\vec{L}_x = L_x \vec{e}_x$ and $\vec{L}_\tau = \tau L_x \vec{e}_x + L_y \vec{e}_y$.
We work in the corresponding Landau gauge such that our orbitals have a well-defined momentum in the direction of $\vec{L}_\tau$.
Finally, we note $\vert m \rangle = c^\dagger_m \vert 0 \rangle$ the state with an electron at momentum $k_y = \frac{2\pi}{L_y} m$, with $m \in [0, N_\phi[$.\\

The computation of the guiding center density is straightforward at the second quantization level. 
Using Eq.~\eqref{eq:CompGC}, we obtain
\begin{equation}
\overline{\rho}(\vec{q}) =  \sum\limits_{n} e^{i\pi \frac{n_x n_y}{N_\phi}}e^{2i\pi \frac{n_x n}{N_\phi}} c^\dagger_{n+n_y\,[N_\phi]} c_n
\end{equation}
where we used the notation $\vec{q} = n_x \left( \frac{2\pi}{L_x} \vec{e}_x + \frac{2\pi \tau}{L_y} \vec{e}_y \right) + n_y \frac{2\pi}{L_y} \vec{e}_y$. 
The expression of $\overline{S}$ follows straighforwardly.
To extract the long-range contributions $\overline{S}_{4/6}$, we rely on finite differences or fits at small $\vec{q}$.
Due to the finite size of the torus, we have access to only a limited subset of momenta, such that the smallest accessible $q$ is inversely proportional to $L_x$ and $L_y$.
To get sufficient data, we need to vary $N_\phi = \frac{L_x L_y}{2 \pi}$, $r$ or $\tau$.
We discuss in the next section the different approaches. \\

Finally, on the torus, even Abelian (and non-topological) states have degeneracies.
As a brief reminder, the algebra of magnetic translations divides the Hilbert space into $N_d \times N_\phi$ symmetry sectors where $N_d = \mathrm{gcd}(N_e, N_\phi)$.
We note the corresponding symmetry sectors $(s, t)$.
The $q$ sectors $(s, t + n N_d)$, $n \in [0, q-1]$ are strictly equivalent, as they are mapped onto each others by global translations.
Due to the degeneracy, we have in principle a choice on how to compute the structure factors.
We show below that all possible choices are equivalent here.
Firstly, the guiding center structure factors are trivially diagonal in momentum space, such that (noting $\vert GS(s, t) \rangle$ the groundstate in the sector $(s, t)$)
\begin{equation}
\langle GS(s, t)\vert \overline{S}(\vec{q}) \vert GS(s', t')\rangle = 0  \text{ if } s\neq s' \text{ or } t \neq t'.
\end{equation}
Secondly, denoting $T$ the operator shifting all orbitals by $1$ magnetic length, 
\begin{equation}
T \overline{\rho}(\vec{q}) T^{-1} = e^{2i\pi \frac{n_x}{N_\phi}} \overline{\rho}(\vec{q}).
\end{equation}
Consequently, upon a global translation of the state, the guiding center structure factors themselves are left unchanged.
Taken together, assuming we have a unique groundstate at $(s_0, t_0)$ (i.e. Abelian excitations), up to the trivial degeneracy, we obtain
\begin{equation}
\langle GS(s_0, t_0)\vert \overline{S}(\vec{q}) \vert GS(s_0, t_0)\rangle \equiv \overline{S}_{s_0, t_0} = \frac{1}{q} \sum\limits_{j = 0}^{q-1} \overline{S}_{s_0, t_0 + j N_d},
\end{equation}
i.e., we can indifferently evaluate the structure factors in a single groundstate, in the translation-invariant density matrix
\begin{equation}
\rho = \frac{1}{q} \sum\limits_{j = 0}^{q-1} \vert GS(s_0, t_0 + j N_d) \rangle \langle GS(s_0, t_0 + j N_d) \vert 
\end{equation}
or in any statistical or quantum superposition of the degenerate groundstates.\\

The case of non-Abelian degeneracies will be discussed in Sec.~\ref{subsec:nonAb}.
 
\subsection{Infinite DMRG on the cylinder}
We use infinite DMRG (iDMRG) to compute the groundstates of the parent Hamiltonians\citep{Haldane1983, Greiter1991, Lee2015, Chen2017, Bandyopadhyay2020} of the Read-Rezayi family.
Our code is based on the \href{https://itensor.github.io/ITensors.jl/stable/index.html}{ITensors.jl}~\cite{itensor} library and its subpackage \href{https://github.com/ITensor/ITensorInfiniteMPS.jl}{ITensorInfiniteMPS.jl}.
We note that it is possible to compute analytically the matrix product state (MPS) representations of all the states that we study in this article\citep{Zaletel2012, Estienne2013, Estienne2013b, Estienne2015}.
As our goal is also to verify that we are able to extract the correct topological information from states obtained numerically, i.e., within the allowed precision on both MPS and Hamiltonians, we chose not to employ this approach.\\

We consider an infinite cylinder periodic in the $L_y$ direction, and implemented both particle and momentum conservation, following the pioneering works of Refs.~\onlinecite{Zaletel2013, Zaletel2015}.
For the infinite cylinder, we denote $N_\phi$ the number of orbitals in the unit cell.
Due to the two $U(1)$ symmetries, two-site DMRG cannot explore the full phase-space.
We use subspace expansion\citep{Hubig2015, tenpy} to alleviate this problem without resorting to expensive larger DMRG steps.
We found that it is advantageous to use random Hamiltonians to perform the subspace expansion given the large bond dimension (up to $2200$) of our models.\\

To obtain the relevant Hamiltonians, we first use a factorized form of the $N-$body pseudopotentials\citep{Haldane1983, Lee2013, Ortiz2013, Papic2014, Kang2017, Repellinpersonal} and compute their translation-invariant second-quantized representation.
The largest Hamiltonians we consider in this work are sums of millions of elementary operators.
To keep their matrix product operator (MPO) representation under check, we compress them at several stages.
We first compute the (quasi-)exact representation of the translation-invariant Hamiltonian as a finite MPO.
We use this MPO to build a corresponding, exact, infinite MPO (iMPO) using a matrix of tensors representation.
Finally, we implemented the iterative compression algorithm proposed in Ref.~\onlinecite{Parker2020}.
To give a concrete order of magnitude, the largest Hamiltonians we consider have of the order of $5$ millions terms larger than $10^{-7}$.
After truncating singular values below floating precision, we obtain an iMPO of bond dimension $\chi_\mathrm{MPO}\approx 2200$.
We are able to obtain a groundstate iMPS with $\chi_\mathrm{MPS} \leq 8192$ whose energy per site is correct at $10^{-5}$ (the bulk gap is of order $1$).
At a fixed bond dimension $\chi_\mathrm{MPS}$, we converge the energy per site and the entropy with a precision better than $10^{-6}$.\\

There is a trade-off for such parent Hamiltonians between $\chi_\mathrm{MPO}$ and $\chi_\mathrm{MPS}$.
On one hand, an approximate iMPS representing the exact Read-Rezayi states\citep{Read1999} has a lower bond dimension than for more generic groundstates at a given precision due to the sparsity and structure of the entanglement spectrum.
On the other hand, approximating the Hamiltonian, either by discarding small coefficients or through the compression scheme we used, is equivalent to introducing a small perturbation.
The true groundstate of the approximate MPO consequently has a non-universal part, and requires a larger $\chi_\mathrm{MPS}$ when $\chi_\mathrm{MPO}$ is reduced.
Given that the non-universal contributions can significantly affect the quantities of interest in the paper, for the more complex states, we pushed the precision of the iMPO.\\

Finally, we take advantage of the infinite length of the cylinder and only evaluate $\overline{S}(q_x, 0)$.
The small $\vec{q}$ behavior is obtained either using numerical fits or
\begin{equation}
\overline{S}_{2n} = \frac{(-1)^n}{N_\phi (2n)!} \sum\limits_{l=1}^{N_\phi} \sum\limits_{m \in \mathbb{Z}}(\frac{2m\pi}{L_y})^{2n} \langle \delta n_l n_{l+m} \rangle.
\end{equation}
The infinite sum can be truncated as the expectation value decays exponentially at large distance\citep{Kumar2023}.
We note that the convergence of these series can be extremely slow with $m$, and can show significant instabilities.

\subsection{Structure factors for non-Abelian states}\label{subsec:nonAb}
Non-Abelian states on the torus and infinite cylinder have non-equivalent, degenerate groundstates.
On the torus, this degeneracy is connected to the non-trivial genus of the manifold.
On the cylinder, the different states correspond to different boundary conditions with non-Abelian excitations at the edges. \\

In the thermodynamic limit, the structure factors should not depend on the state under consideration, at least at finite momenta.
Indeed, the different groundstates are not locally distinguishable.
Nonetheless, for finite systems, there is no reason for these different states to have the same structure factors.
The conjectures for $\overline{S}_4$ and $\overline{S}_6$ do not consider this topological degeneracy, as the coefficients are evaluated in the thermodynamic limit only.
Note that considering a sphere, where non-Abelian models are non-degenerate, does not solve the question of the finite-size effects and strongly limits the accessible momenta. \\

For both manifolds under consideration, we propose a simple recipe: we should average the structure factors over the different topological sectors.
Indeed, on a finite torus, by analogy to the computation of topological invariants, we can compute the structure factors on the (normalized) projector to the groundstate manifold.
We verify below that it does indeed reduce finite-size oscillations.
On the infinite cylinder, it is similarly natural to average over all the different ways to split an infinite, periodic state with zero anyonic total charge in a non-periodic state with open boundary conditions.
To give a more concrete example, let us consider the Moore-Read state, i.e., a non-Abelian model where the local excitations are Majorana fermions (denoted $\gamma$).
The Moore-Read state can be understood as a gapped $p$-wave unconventional superconductor\citep{Read2009} composed of pairs of coupled Majorana fermions in the bulk.
When splitting the system, we either cut the state between two uncoupled pairs, leading to standard open boundary conditions $1 \times 1$, or cut through a pair, such that the boundary conditions are $\gamma \times \gamma$.
In the latter case, a single Majorana excitation is present at each extremity.
The relevant structure factors will then be the average of the two possible boundary conditions.\\

The Moore-Read Pfaffian is studied in more details in Sec.~\ref{subsec:MR}.
We extend non-trivially our formula to richer non-Abelian states in Secs.~\ref{subsec:RR35} and ~\ref{subsec:RR46}).

\section{Structure factors in the Read-Rezayi series}\label{sec:SFRR}
\begin{figure}
\begin{center}
\includegraphics[width = \linewidth]{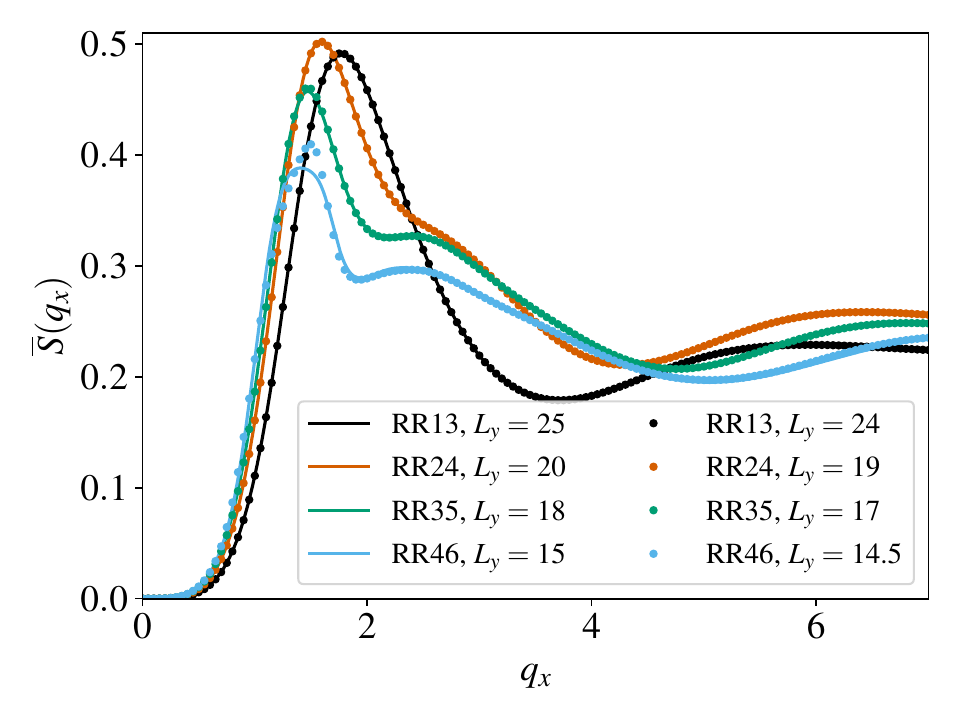}
\end{center}
\caption{Structure factors for the Read-Rezayi family of states for different cylinder width $L_y$. The Laughlin (RR13), Pfaffian (RR24) and Read-Rezayi $3/5$ states are well-converged with $L_y$, while significant deviations are still visible for the Read-Rezayi $4/6$ state. The width is nonetheless sufficient to extract the long-range contributions to the structure factors with a good accuracy.}
\label{fig:AllGCSF}
\end{figure}

The $\mathbb{Z}_k$ Read-Rezayi family\citep{Read1999} is a sequence of FQH states which sustains non-Abelian topological order for $k>1$.
In this article, we focus on their fermionic versions, whose filling is $\nu = \frac{k}{k+2}$ for $k \in \mathbb{N}^*$.
They are the groundstates of the simplest non-trivial, momentum-conserving repulsive pseudopotentials at $k+1$ particles, given for fermions by
\begin{equation}
V_{k} =  \prod\limits_{j=1}^k \nabla^{2j}_{\vec{r}_j}  \prod\limits_{j=1}^k \delta(\vec{r}_{k+1} - \vec{r}_j)
\end{equation}
The root configurations\citep{Li2008} of the corresponding groundstates follow the simple rule of "no more than $k$ particles in $k+2$ consecutive orbitals".
The $k=1$ state is nothing but the Laughlin\citep{Laughlin1983} state at filling $1/3$.
The $k=2$ state is the Moore-Read Pfaffian\citep{Moore1991} whose quasiparticles are non-Abelian Majorana fermions.
The quasiexcitations for $k=3$ (resp. $k=4$) are the $\mathbb{Z}_3$ (resp. $\mathbb{Z}_4$) parafermions.
All these states have topological spin $s=1$.
The edge central charge and the degeneracy of these states on the torus and infinite cylinder are listed in Table.~\ref{tab:PropertiesRR}.
The summary of our numerical results are also indicated in the Table, while the different interpolation methods are discussed in the rest of the paper.
Finally, representative guiding center structure factors of the different states are displayed in Fig.~\ref{fig:AllGCSF}.
In particular, we are able to obtain high-precision estimates of $\overline{S}_4$ and $\overline{S}_6$ for the Read-Rezayi states at filling $\nu = 4/6$ even though the function is not fully converged with $L_y$ at intermediate distances.

\begin{table}
\newcolumntype{N}{@{}m{0pt}@{}}
\begin{tabular}{|c|c|c|c|c|c|c|c|c|c|c|N}
\hline $k$ & $\nu$ & $c_0$ & $D_T$ & $D_\mathrm{cyl}$ & $\overline{S}_4^\mathrm{theo}$ & $\vert \delta \overline{S}_4 \vert $ & $\delta \overline{S}_4/\overline{S}_4$& $\overline{S}_6^\mathrm{theo}$ & $\vert \delta \overline{S}_6 \vert $ & $\delta \overline{S}_6/\overline{S}_6$ &\\[2pt] \hline 
& & & & & & & & & & &\\[-0.8em]
$1$ & $\frac{1}{3}$ & $0$ & $3 \times 1$ & $1$ & $\frac{1}{12}$ & $<1e^{-4}$ & $<0.1\% $ & $\frac{5}{144}$ & $5e^{-5}$ & $ 0.15\%$ & \\[3pt] \hline
& & & & & & & & & & &\\[-0.8em]
$2$ & $\frac{1}{2}$ & $\frac{1}{2}$ & $2 \times 3$ & $2$& $\frac{1}{8}$ & $6e^{-4}$ & $0.5\% $ & $\frac{5}{96}$ & $1e^{-4}$ & $ 0.2\%$ &\\[3pt] \hline
& & & & & & & & & & &\\[-0.8em]
$3$ & $\frac{3}{5}$ & $\frac{4}{5}$ & $5 \times 2$ & $2$ & $\frac{3}{20}$ & $8e^{-4}$ & $0.5\% $ & $\frac{1}{16}$ & $2e^{-4}$ & $ 0.3\%$&\\[3pt] \hline
& & & & & & & & & & &\\[-0.8em]
$4$ & $\frac{2}{3}$ & $1$ & $3 \times 4$  & $3$ & $\frac{1}{6}$ & $8e^{-3}$ & $5\% $ & $\frac{5}{72}$ & $4e^{-4}$ & $ 5\%$ &\\[3pt] \hline
\end{tabular}
\caption{Properties of the Read-Rezayi state $k$ at filling $\nu$. The total central charge for the edge modes is given by $c = 1 + c_0$, where $c_0$ is the contribution of the chargeless modes. $D_T$ is the degeneracy of the groundstate on a finite torus, expressed as the trivial degeneracy times the non-Abelian degeneracy. $D_\mathrm{cyl}$ counts the number of non-equivalent states that appear on an infinite cylinder. It corresponds to the number of non-equivalent root configurations. The last four columns summarize our numerical results, listing the theoretical expected values of $\overline{S}_{4/6}$ and the best absolute and relative precision obtained using our numerical methods. For reference, for all these models, a precision of $0.1$ on the extracted central charge correspond to a relative precision of $\approx 1.5\%$ on $\overline{S}_6$.}
\label{tab:PropertiesRR}
\end{table}

\subsection{Laughlin $\nu = 1/3$}
This section briefly studies the guiding center structure factors of the Laughlin\citep{Laughlin1983} state at filling $\frac{1}{3}$, as a benchmark.
It is an Abelian state and therefore does not present any topological degeneracy.
On the infinite cylinder, we work with a unit-cell of size $3$ and it is sufficient to consider the groundstate obtained using the starting root configuration $100$.\\

\begin{figure}
\begin{center}
\includegraphics[width = \linewidth]{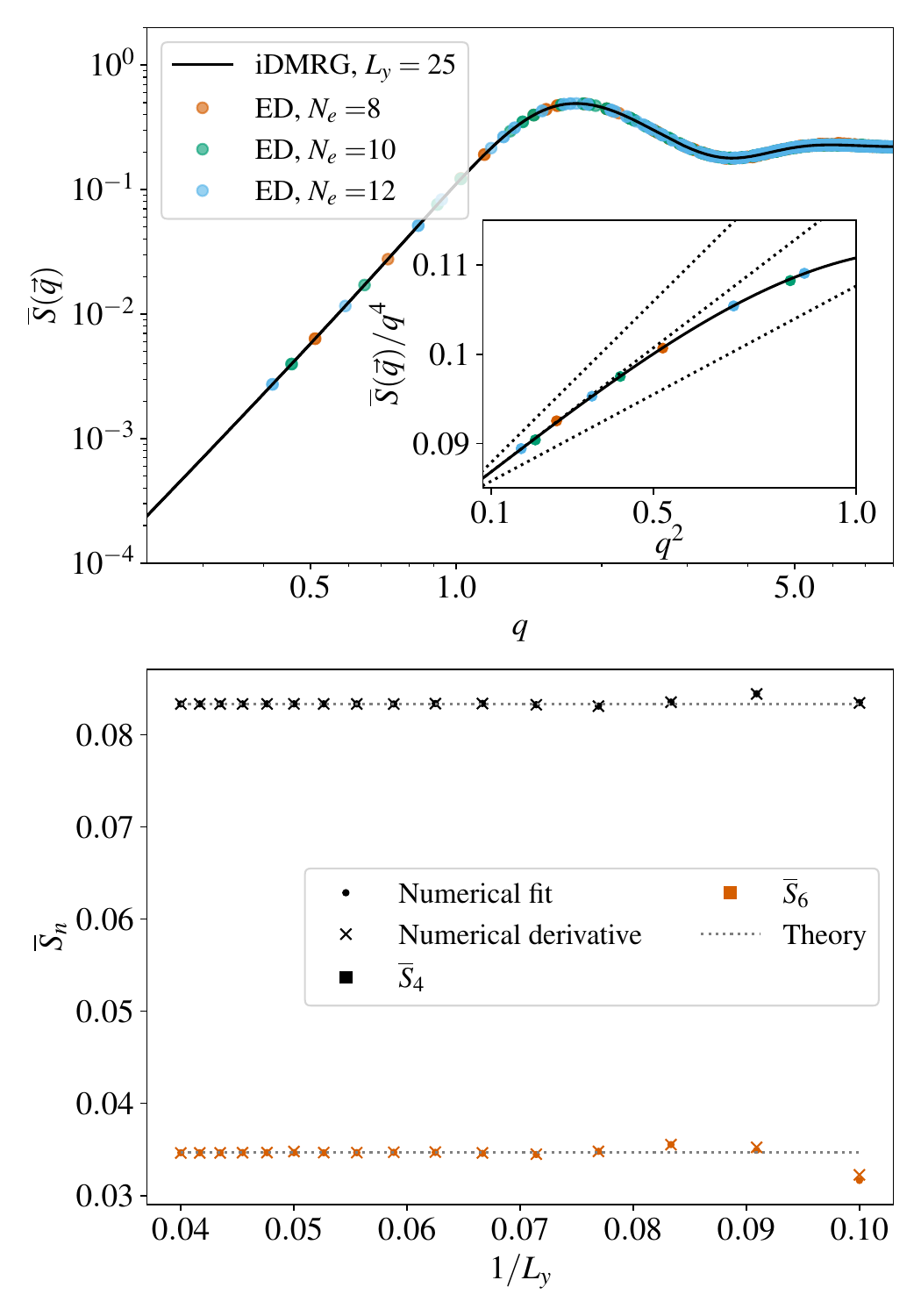}
\end{center}
\caption{Guiding centers for the Laughlin state. Top: Structure factors obtained from iDMRG for $L_y = 25$, and from ED for several values of $N_e$, $r=1.0$ and $\tau = 0$. We observe already a very good agreement. In inset, we zoom on small $\vec{q}$. The dotted lines are guides to the eyes showing the predicted curves using only the theoretical $\overline{S}_4$ and $\overline{S}_6$, with $c$ taken to be $0$, $1$ and $2$. Bottom: $\overline{S}_4$ and $\overline{S}_6$ obtained from iDMRG for $L_y$ ranging from $10$ to $25$. We observe a quick convergence towards the theoretical values.}
\label{fig:SFLL13}
\end{figure}

Our numerical results are summarized in Fig.~\ref{fig:SFLL13}.
For this state, it has been proved that
\begin{equation}
\overline{S}_4 = \frac{1}{12}, \qquad \overline{S}_6 =\frac{5}{144}.
\end{equation}
Varying the number of electrons from $N_e = 8$ to $12$ and the aspect ratio from $0.8$ to $1$, $\overline{S}_4$ can already be obtained with a precision better than $1\%$ and $\overline{S}_6$ within $5\%$.

Computations on the infinite cylinder give more precise results.
For the largest system sizes we study ($L_y = 25$), the relative error on $\overline{S}_4$ is of order $5\times 10^{-5}$.
The relative error on $\overline{S}_6$ is of order $2\times 10^{-3}$.

\subsection{Moore-Read Pfaffian state}\label{subsec:MR}
The second state in the Read-Rezayi series is the Moore-Read\citep{Moore1991} Pfaffian at filling $\nu = \frac{2}{4}$.
It is characterized by the presence of Majorana excitations.\\

We start by discussing exact diagonalization on the torus.
With $N_\phi = 0 \text{ mod } 4 $, it has three non-equivalent groundstates in the sectors $(N_d/2, n N_d)$, $(0, (n + \frac{1}{2}) N_d)$ and  $(N_d/2, (n + \frac{1}{2}) N_d)$.
As these states belong to different symmetry sectors, the discussion on degeneracy in Sec.~\ref{sec:ED} is still valid.
Before considering the effect of the degeneracy, we investigated the systematic finite-size effects arising from the different ways to evaluate the small $q$ behavior of the structure factors.
As discussed previously, we can vary the size, the aspect ratio, or the boundary twist of our torus to have access to more momenta. 
Varying the system sizes introduce larger finite-size effects.
Varying the aspect ratio or the twist angle breaks the rotation invariance at large distances.
We found that focusing on the largest systems ($N_\phi = 32$ here) and varying the twist angle leads to the best estimations of $\overline{S}_4$ and $\overline{S}_6$.
Interestingly, significant deviations to the theoretical predictions are observed for the smallest momenta available, which we naturally interpret as an effect of the boundary.\\

Our method fixed, we return to the question of the groundstate degeneracy.
Because all groundstates belong to different symmetry sectors, it is equivalent to consider statistical mixing or quantum superpositions of the different groundstates.
The oscillations around the theoretical prediction are strongly reduced when we average the structure factors over the different groundstates, confirming the intuition of Sec.~\ref{subsec:nonAb}.
The relevant numerical data are available in Appendix.~\ref{app:FS-torus}.\\

\begin{figure}
\begin{center}
\includegraphics[width=\linewidth]{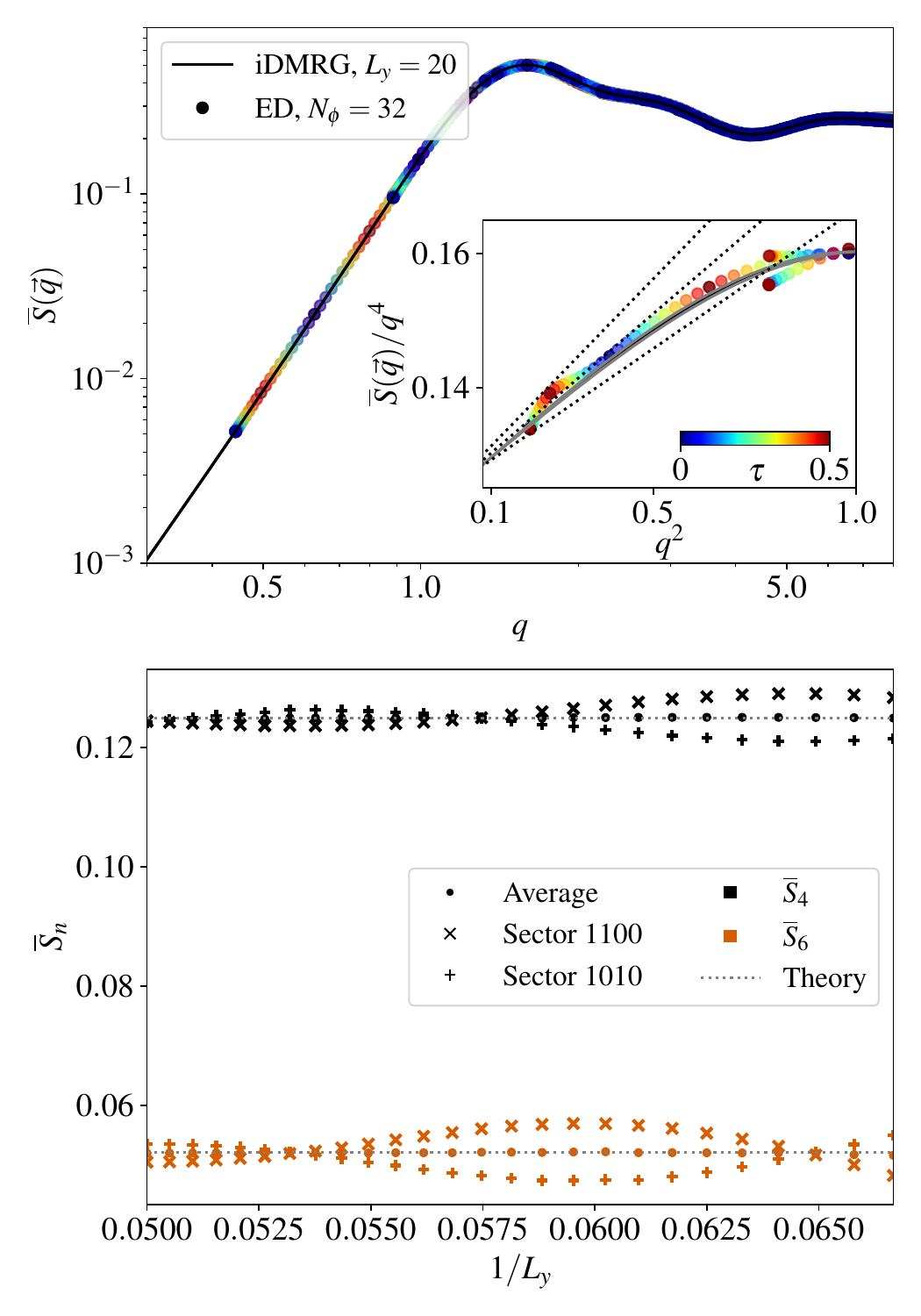}
\end{center}
\caption{Top: averaged structure factors in the Moore-Read state on an infinite cylinder with $L_y = 20$ (black line) and on a finite torus with $N_\phi = 32$ and $r=1.0$ (dots). In inset, we represent $\overline{S}(q)/q^4$ at small $q$. The two gray lines are the individual contributions of the two groundstates on the cylinder, while the dotted lines are guides to the eyes showing the predicted curves using only the theoretical $\overline{S}_4$ and $\overline{S}_6$, with $c$ taken to be $\frac{1}{2}$, $\frac{3}{2}$ and $\frac{5}{2}$. Bottom: $\overline{S}_4$ and $\overline{S}_6$ extracted from iDMRG as a function of $L_y$. Even at $L_y > 20$, we observe exponentially-decaying oscillations in each sectors, . Remarkably, averaging over the two groundstates nearly entirely cancels these oscillations, such that the relative error for both coefficients is below $1\%$.
}
\label{fig:GCSF-MR}
\end{figure}

Finally, we show the structure factors averaged over all states for $N_\phi = 32$ and $\tau = 1$ in Fig.~\ref{fig:GCSF-MR}.
After a careful interpolation, we are able to extract the correct $\overline{S}_4$ and $\overline{S}_6$ within a few percent.\\

On the infinite cylinder, we perform a similar analysis.
The two root configurations $1100$ and $1010$ correspond to the two sectors of the $\mathbb{Z}_2$ Majorana fermions, as detailed in Sec.~\ref{subsec:nonAb}.
In Fig.~\ref{fig:GCSF-MR}, we show the structure factors for both groundstates for different values of $L_y$.
In both states, $\overline{S}_4$ and $\overline{S}_6$ appear to converge exponentially towards the theoretical prediction on the sphere.
In fact, we can extract very accurate estimates of these two coefficients with a simple naive fit of the form:
\begin{equation}
\overline{S}_n = \overline{S}_n(\infty) + A e^{-\frac{L_y}{\xi}} \cos(\omega L_y + \phi). \label{eq:Fit}
\end{equation}
Note that it should not come as a surprise: in the strict thermodynamic limit, the two groundstates should be locally indistinguishable and also locally indistinguishable from the wavefunctions on the sphere.
We should therefore recover the predicted scaling of $\overline{S}(\vec{q})$ at least for intermediate distances.

Despite the relatively large cylinder width, the variations around the theoretical values remain significant.
Following our proposed recipe in Sec.~\ref{subsec:nonAb}, we should average the structure factors over the different topological sectors.
The root configuration $1100$ on the infinite cylinder can be understood as the Pfaffian state with trivial boundary conditions.
Conversely, the configuration $1010$ corresponds to a state with edge Majorana excitations.
The entropy difference\citep{Kitaev2006, Levin2006, Zozulya2007, Estienne2015} between the two states indeed converges to $\log \sqrt{2}$ in the thermodynamic limit (see App.~\ref{app:TopoEnt}).
We therefore compute the averaged structure factor $\overline{S}_\mathrm{av} = \frac{1}{2}\left( \overline{S}_{1100} + \overline{S}_{1010} \right)$ which shows a remarkable suppression of the oscillations.
We obtain $\overline{S}_4$ and $\overline{S}_6$ with a precision better than $1\%$, without any fit.

\subsection{The Read-Rezayi $3/5$ state}\label{subsec:RR35}
The third state in the Read-Rezayi series at $\nu = \frac{3}{5}$  admits $\mathbb{Z}_3$ non-Abelian parafermionic excitations.\\

\begin{figure}[ht]
\begin{center}
\includegraphics[width=\linewidth]{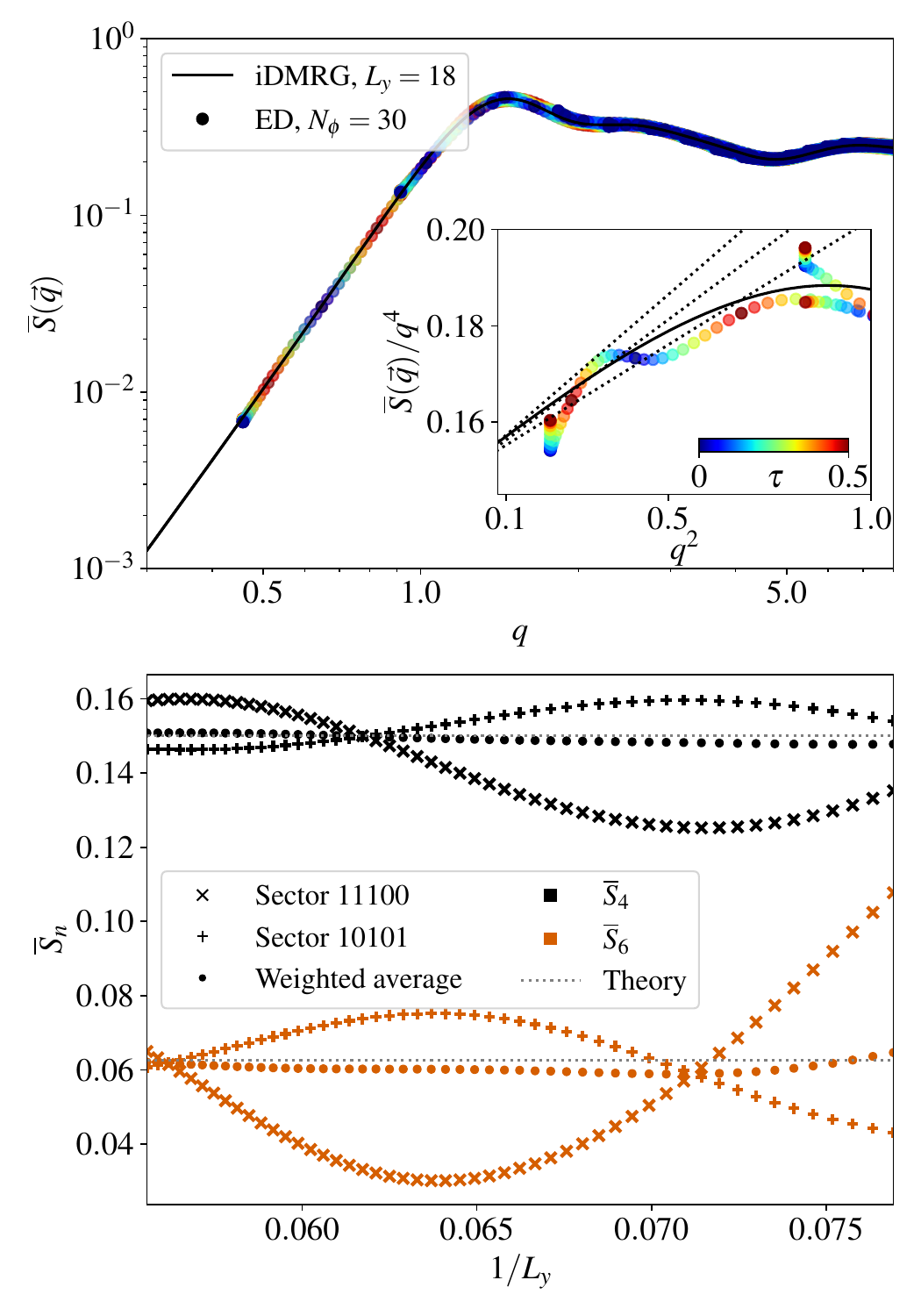}
\end{center}
\caption{Top: averaged structure factors for the Read-Rezayi state $\nu = \frac{3}{5}$ on an infinite cylinder with $L_y = 18$ (black line) and on a finite torus with $N_\phi = 30$ and $r=1.0$ (dots). In inset, we represent $\overline{S}(q)/q^4$ at small $q$. The dotted lines indicates the small $q$ behavior for $c =-\frac{1}{5}$, $\frac{4}{5}$ and $\frac{9}{5}$. Bottom: $\overline{S}_4$ and $\overline{S}_6$ extracted from iDMRG as a function of $L_y$. While large macroscopic oscillations can be seen for the individual groundstates, their weighted average shows remarkably small boundary effects. 
}
\label{fig:GCSF-RR35}
\end{figure}

Starting on the torus, in addition to the trivial $5$-fold degeneracy, it admits $2$ groundstates in the sector $(0, 0)$ for $N_d$ odd and $(N_d/2, N_d/2)$ for $N_d$ even for a total of $10$ groundstates.
As the groundstates are in the same momenta sectors, the structure factors now depend in principle of the chosen  superposition of groundstates.
Given our previous results on the Pfaffian, and by analogy with the standard computations of topological invariants, we work with the equal-weight density matrix 
\begin{equation}
\rho_{\frac{3}{5}} = \frac{1}{2} \left(\vert GS_1 \rangle \langle GS_1 \vert + \vert GS_2 \rangle \langle GS_2 \vert \right)
\end{equation}
describing the groundstate manifold.
The corresponding structure factors are simply the average of the structure factors obtained for the two orthogonal groundstates.
They are shown in Fig.~\ref{fig:GCSF-RR35} for $N_\phi = 30$ and different twist angles.
We are able to extract a relatively accurate prediction for $\overline{S}_4$ (within $\approx 5\%$ of its theoretical value), and therefore we can reliably extract the topological spin.
The strong finite size effects prevent us from concluding on $\overline{S}_6$.
Note that the main limiting factor for ED computations here is not the dimension of the Hilbert state, but the lack of sparsity of the Hamiltonians.
Due to this lack of sparsity, the standard matrixless iterative algorithms to compute eigenvalues are largely inefficient.

We now turn to the infinite cylinder.
There are now two inequivalent root configurations $11100$ and $11010$.
The root configuration $11100$ corresponds to a state with trivial boundary conditions while $11010$ can be interpreted as having edge parafermionic excitations.
Using appropriate starting states, we are able to converge to the two groundstates.
We verified convergence by measuring the difference in entropy in the thermodynamic limit $\delta \gamma = \log \phi$, with $\phi$ the golden ratio\citep{Estienne2015} (see App.~\ref{app:TopoEnt} for more details).
In Fig.~\ref{fig:GCSF-RR35}, we show the structure factors for both groundstates for different values of $L_y$.
The structure factors still converge exponentially towards their theoretical values.
The finite-size effects are significantly larger than for the Moore-Read states due to the larger correlation lengths.
A naive fit of the oscillations using Eq.~\eqref{eq:Fit} is also enough to get $\overline{S}_4$ within $1\%$ and $\overline{S}_6$ within $\approx 2 \%$. \\

Finally, we verify non-trivially our recipe for non-Abelian states.
As the excitations are composed of $\mathbb{Z}_3$ parafermions, the possible boundary configurations are now $1 \times 1$, $\tau \times \bar{\tau}$ and $\bar{\tau} \times \tau$ ($\tau$ denotes a $\mathbb{Z}_3$ parafermions and $\overline{\tau}$ is the corresponding anti-particle).
The two configurations $\tau \times \bar{\tau}$ and $\bar{\tau} \times \tau$ have identical contributions to the structure factors (both due to inversion symmetry and due to the structure of the $\mathbb{Z}_3$ parafermionic space).
The root configuration $11100$ corresponds to the trivial boundary conditions, while $11010$ corresponds to $\tau \times \bar{\tau}$.
The weighted-average structure factor to study is therefore
\begin{equation}
 \overline{S}_{\frac{3}{5}} = \frac{1}{3}\left( \overline{S}_{11100} + 2 \overline{S}_{11010} \right).\label{eq:avSFRR35}
\end{equation}
Using $ \overline{S}_{\frac{3}{5}}$, we are able to verify the conjectured formula for $\overline{S}_4$ and $\overline{S}_6$ within $1\%$ even for relatively small cylinder of size $L_y = 18.0$ without any fitting.

\subsection{The Read-Rezayi $4/6$ state}\label{subsec:RR46}
Finally, the fourth state in the Read-Rezayi series at $\nu = \frac{4}{6}$ admits $\mathbb{Z}_4$ non-Abelian parafermionic excitations.
It is the groundstate of a $5$-body parent Hamiltonian.
On the torus, the lack of sparsity of the Hamiltonian largely prevents us from reaching relevant system sizes.
The topological spin can still be evaluated using $\overline{S}_4$ if we assume that the formula Eq.~\eqref{eq:defS4} is correct, but $\overline{S}_6$ is out of reach.\\

We therefore focus on the infinite cylinder.
The complexity of the Hamiltonian still limits us to relatively small cylinders, as at $L_y = 15.0$, the translation invariant Hamiltonian already includes $\approx 5 \times 10^6$ terms larger than $10^{-7}$.
It is nonetheless enough to obtain good estimates of the structure factors.
There are now three distinct root configurations that satisfy the minimal constraints: $111100$ (trivial boundary conditions), $111010$ (a $\mathrm{Z}_4$ parafermion and its antiparticle) and $110110$ (two $\mathrm{Z}_4$ parafermions at each extremities).
Both excitations lead to significantly higher entanglement (quantum dimension $\sqrt{3}$ and $2$).
The configuration $110110$ is especially challenging to obtain numerically: the local momenta can be identical to $111100$'s, while its entanglement entropy is significantly higher.
Simulations therefore tend to relax to the root configuration $111100$ if we allow large changes in the wavefunctions at each iDMRG iteration.
In practice, we verify that we obtain the correct groundstates by computing the difference in topological entanglement entropy, which appear to converge towards $\log \sqrt{3}$ and $\log 2$ in the thermodynamic limit.
We refer to App.~\ref{app:TopoEnt} for the numerical data.\\

\begin{figure}[ht]
\begin{center}
\includegraphics[width=\linewidth]{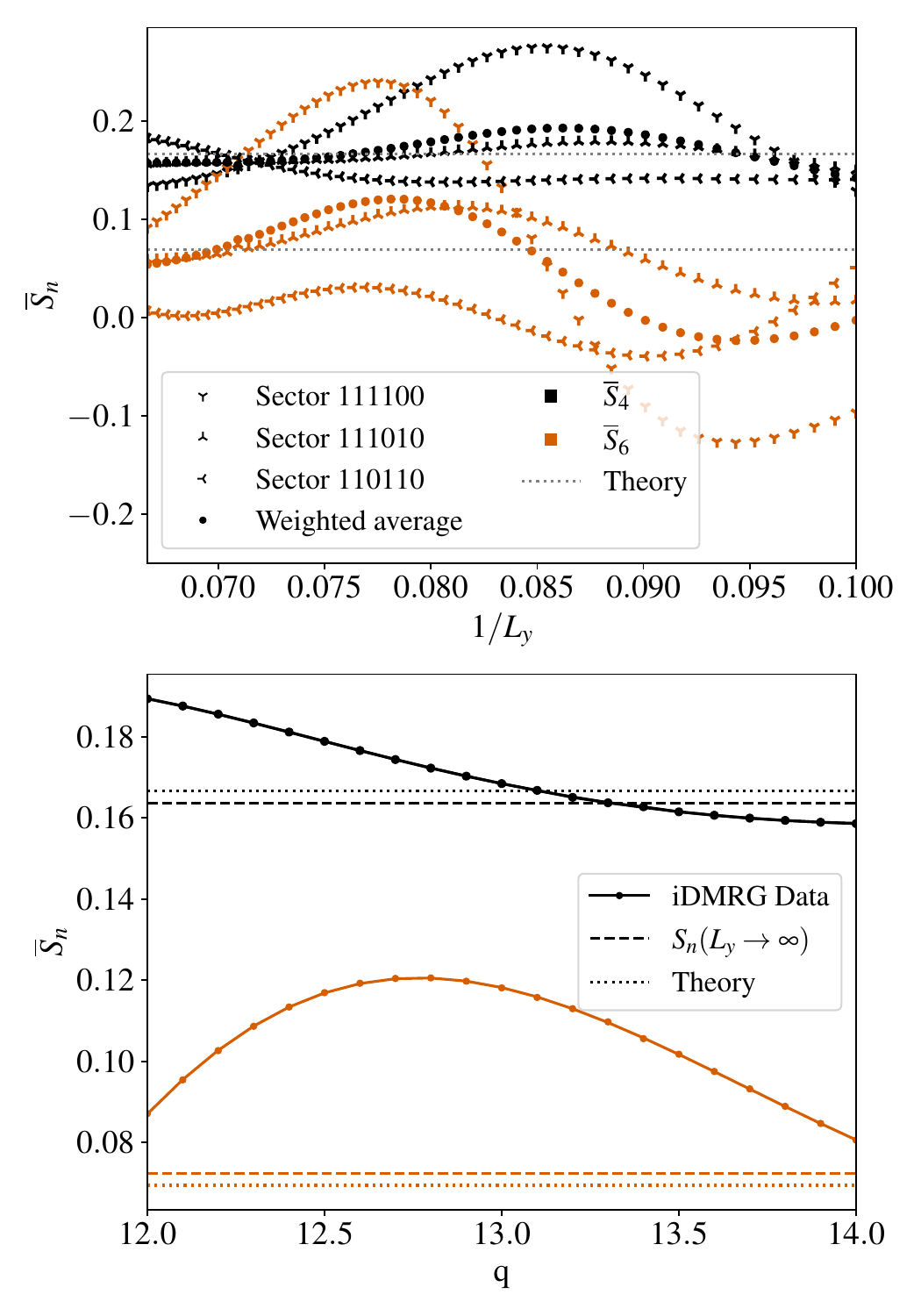}
\end{center}
\caption{Top: $\overline{S}_4$ and $\overline{S}_6$ as a function of $L_y$ in the Read-Rezayi state at $\nu = \frac{4}{6}$. The large oscillations are reduced once the correct superposition is chosen. Bottom: we can also fit the averaged contributions with exponentially decaying oscillations. 
}
\label{fig:GCSF-RR46}
\end{figure}

In Fig.~\ref{fig:GCSF-RR46}, we show the structure factors for the three groundstates for different values of $L_y$.
Following our recipe, the oscillations are largely suppressed if we consider the averaged structure factor 
\begin{equation}
 \overline{S}_\mathrm{RR46} = \frac{1}{4}\left( \overline{S}_{111100} + 2 \overline{S}_{111010}  + \overline{S}_{110110} \right).\label{eq:avSFRR46}
\end{equation} 
Due to the small $L_y$ and the strong finite-size effects, we still need to proceed to a numerical fit of $\overline{S}_\mathrm{RR46}$ using Eq.~\eqref{eq:Fit} to extract reliable estimates of $\overline{S}_4$ and $\overline{S}_6$.
Note that this fit is here possible because we are in a regime where the main source of error on $\overline{S}_\mathrm{RR46}$ is the finite width.
If we increase the system size, the effect of the iMPO truncation dominates and $\overline{S}_\mathrm{RR46}$ deviates from the theoretical predictions.
Nonetheless, we observe a similar phenomenology as in the previous examples, with a decent agreement between theoretically expected and extracted coefficients (within $\approx 5\%$).

\section{Conclusions}
In this paper, we have shown that the conjectured formulas connecting the first coefficients $\overline{S}_4$ and $\overline{S}_6$ of the guiding center structure factors to the topological spin and the central charge of the edge theory are valid for conformal wavefunctions beyond the Laughlin and Moore-Read states.
The finite system sizes accessible by diagonalization are enough to evaluate these coefficients for the Laughlin or the Moore-Read wavefunctions.
Yet they are insufficient to tackle more complex topological orders due to the larger correlation lengths.
We showed that states obtained from numerical infinite MPOs can be used to verify the conjecture for the Read-Rezayi states at $\nu = \frac{3}{5}$ and $\nu = \frac{4}{6}$.
We are able to verify this conjecture within a few percent, even given the approximations and the relatively small cylinders we are limited to.\\

On the way, we have shown that to efficiently recover the thermodynamic, bulk predictions, one needs to average over the different topological sectors.
In particular, on the torus, we need to average the structure factors over the topologically degenerate groundstates.
Moreover, the optimal strategy appears to be to use different boundary conditions to probe different sublattices in momentum space.
Nonetheless, the smaller momenta accessible show systematic deviations from the thermodynamic prediction due to boundary effects.
MPS on an infinite cylinder allow us to reach larger systems, though the circumference of the cylinder remains limited.
The structure factors predicted in the thermodynamic limit on a sphere or an infinite torus do not take into account the degeneracy induced by the open boundary conditions.
For finite circumferences, we showed that finite-size effects were strongly reduced if we average observables over the different boundary conditions compatible with zero total anyonic charges.\\

Beyond the Read-Rezayi series, these results show that it is possible to obtain the long-distance behavior of the guiding center structure factors in numerical wavefunctions.
In particular, they allow us to distinguish between gapped and gapless states.
In the latter, the dominant contributions to the structure factors generally scale as $k^2$ instead of $k^4$.
Moreover, for gapped states, the inequalities verified by $\overline{S}_4$ and conjectured for $\overline{S}_6$ translate into bounds on the possible topological order. 
This should prove useful for the studies of more complex model Hamiltonians of the FQHE where the characterization of topological order proves challenging.

\begin{acknowledgments}
We thank D. Haldane for introducing us to the physics of structure factors, and N. Regnault and C. Repellin for useful discussions. L.H. acknowledges M. Fishmann and J. Reimers for their help in updating ITensorInfiniteMPS. This work has been supported by the Swiss National Science Foundation (FM) Grant No. 182179.
\end{acknowledgments}

\bibliographystyle{apsrev4-1}
\bibliography{FQHE}

\appendix
\section{Finite-size effects on the torus for the Moore-Read state}\label{app:FS-torus}

\begin{figure}
\vspace*{0.3cm}
\includegraphics[width = \linewidth]{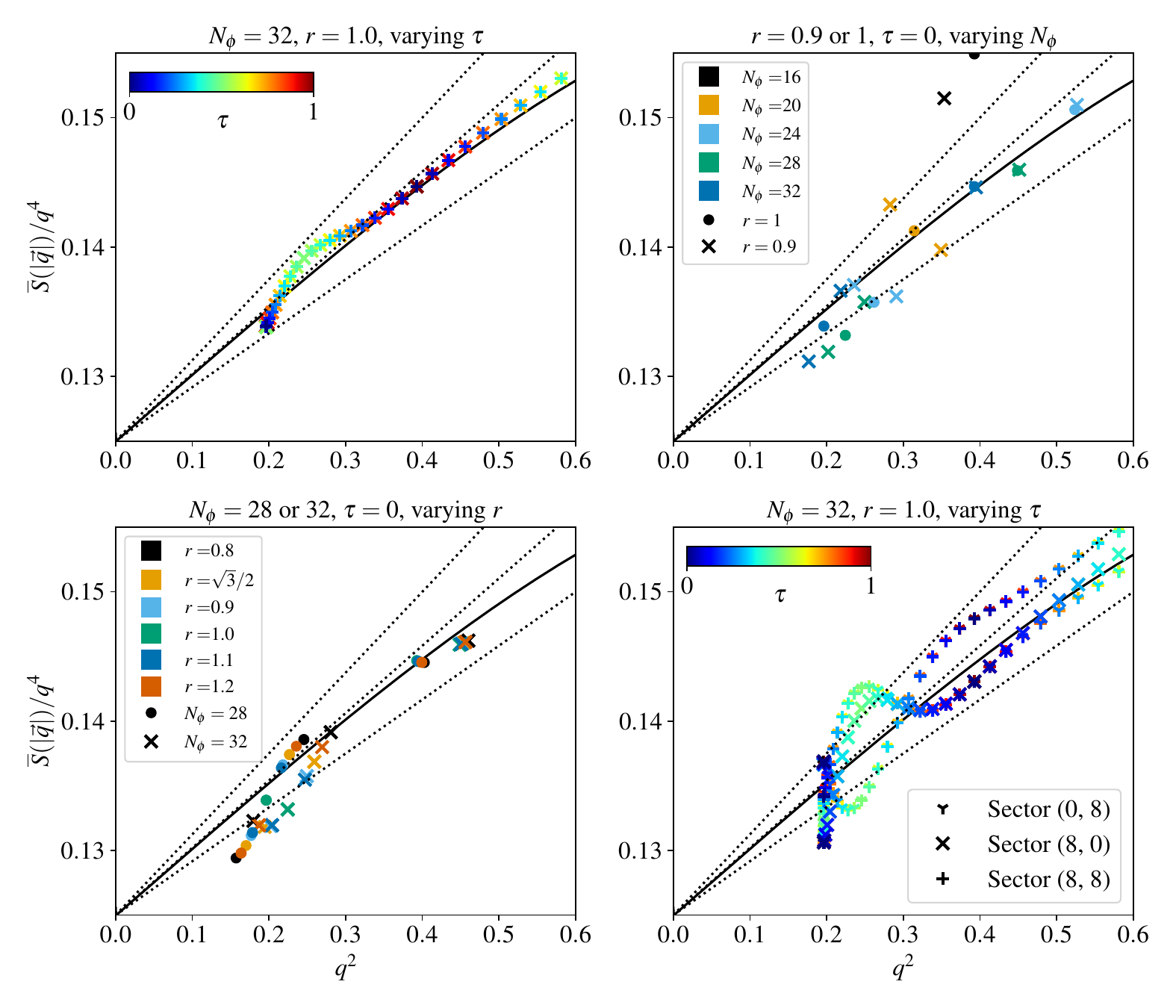}
\caption{Structure factors and their scalings obtained through different variations of the torus. Dots represent data obtained with ED. The black line, included for reference, is the average structure factors obtained from an iDMRG simulation with $L_y=20.0$. The dotted lines are guides to the eyes showing the predicted curves using only the theoretical $\overline{S}_4$ and $\overline{S}_6$, with $c$ taken to be $\frac{1}{2}$, $\frac{3}{2}$ and $\frac{5}{2}$. Top left: we vary the twist angle for $N_\phi = 32$, $r=1$. Top right: we vary the system size $N_\phi$. Bottom left: we vary the aspect ratio of the torus. In all cases, we observe significant finite size effects, though the twisted data is more convenient. In particular, the smaller accessible momenta has strong boundaries effect. Bottom right: instead of averaging over groundstates, we can compute the structure factors separately in each sector. We observe larger oscillations around the theoretical values.}
\label{fig:CompMR}
\end{figure}

We consider the Moore-Read state on a finite torus.
With $N_\phi = 0 \,[4]$, there are three non-equivalent groundstates in the sectors $(N_d/2, n N_d)$, $(0, (n + \frac{1}{2}) N_d)$ and  $(N_d/2, (n + \frac{1}{2}) N_d)$.
Before considering the effect of the degeneracy, we investigate the systematic finite-size effects arising from the different ways to evaluate the small $\vert \vec{q} \vert$ behavior of the structure factors.
As discussed previously, we can vary the size, the aspect ratio, or the boundary twist of our torus to have access to more momenta. 
In Fig.~\ref{fig:CompMR}, we compare these different ways to obtain the structure factors (averaged over the three sectors) and the corresponding small $\vert \vec{q} \vert$ behavior for the Moore-Read state with up to $N_\phi = 32$ orbitals.
For $r=1$, it corresponds to a square torus of length $L \approx 14.2 \gg \xi_{Pf}$ the correlation length of the Pfaffian.
The structure factors are, as expected, mostly a function of $\vert \vec{k} \vert$ only, up to some finite size effects.
The $\overline{S}(\infty)$ is approximately correct in all cases, but the best estimation of $\overline{S}_4$ and $\overline{S}_6$ are obtained by varying the twist.
It should not come as a surprise: in that case, we do not need to consider smaller systems or to explicitely break rotation invariance to obtain our data.
It is interesting to note the significant finite-size deviation at the lowest available momentum.
We naturally interpret it as the effect of the boundaries.
The variations of $\vec{S}(\vec{q})$ at the smallest momenta when varying $\tau$ or $r$ are connected to the significant fluctuations of the Hall velocity in $(r, \tau)$ space at system sizes accessible to ED.

Our method fixed, we return to the question of the groundstate degeneracy.
Because all groundstates belong to different symmetry sectors, it is equivalent to consider statistical mixing or quantum superpositions of the different groundstates.
In Fig.~\ref{fig:CompMR}, we compare the translation invariant contributions of the different groundstates 
More precisely, we compare
\begin{equation}
\overline{S}_{s, t} = \frac{1}{q}\sum\limits_{n = 0}^{q-1} \overline{S}_{s, t + n N_d} 
\end{equation}
to the global average $\overline{S}$.
We see significant deviations and oscillations from the global contributions even at larger system sizes, though they appear to decay slowly with $N_\phi$.
This confirms our intuition that the predictions in the bulk in the thermodynamic limit are best approximated by averaging the structure factors over all groundstates.

\section{Topological entanglement entropy for the Read-Rezayi series}\label{app:TopoEnt}
In a pure state, the entanglement entropy $\mathcal{S}(\mathcal{A})$ quantifies the quantum entanglement between a subsystem $\mathcal{A}$ and the rest of the system.
Let  $\Ket{\psi}$ be a state in the Hilbert space $H = H_\mathcal{A} \otimes  H_{\overline{\mathcal{A}}} $ where $H_\mathcal{A}$ is the Hilbert space of the degrees of freedom in $\mathcal{A}$. 
Its Schmidt decomposition is 
\begin{equation}
\Ket{\psi} = \sum\limits_j \lambda_j \Ket{\alpha_j} \otimes \Ket{\overline{\alpha}_j},
\end{equation}
where $\Ket{\alpha_j} \in H_\mathcal{A}$ and $\Ket{\overline{\alpha}_j} \in H_{\overline{\mathcal{A}}}$.
Its von Neumann entanglement entropy is given by
\begin{equation}
S^\mathrm{vN}(\mathcal{A}) = - \sum\limits_j \lambda_j^2 \log \lambda_j^2.
\end{equation}

\begin{figure}
\begin{center}
\includegraphics[width=\linewidth]{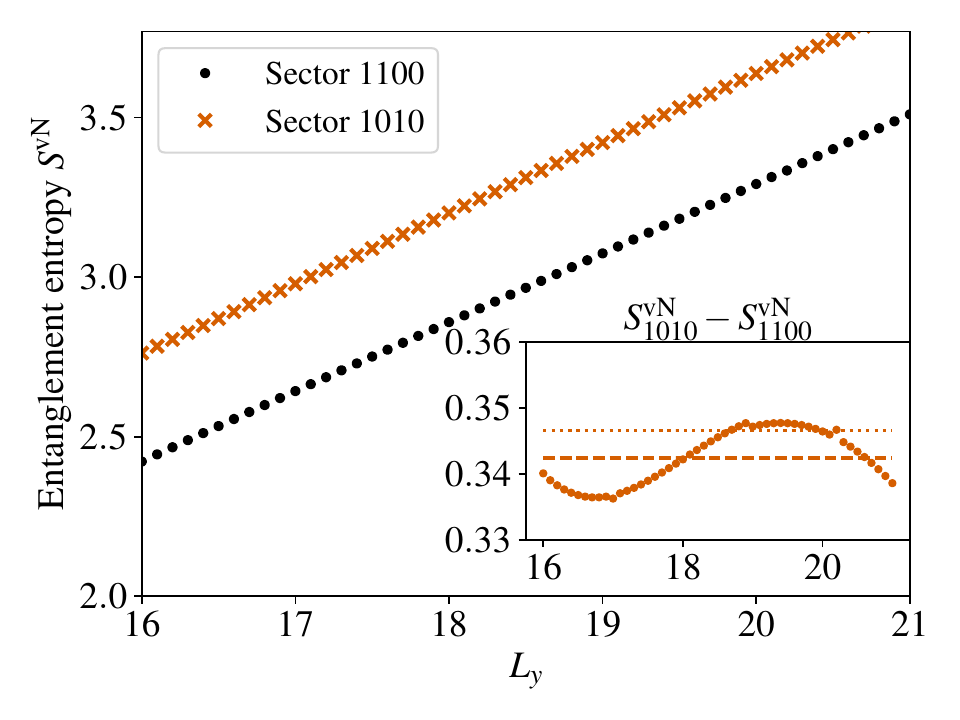}
\includegraphics[width=\linewidth]{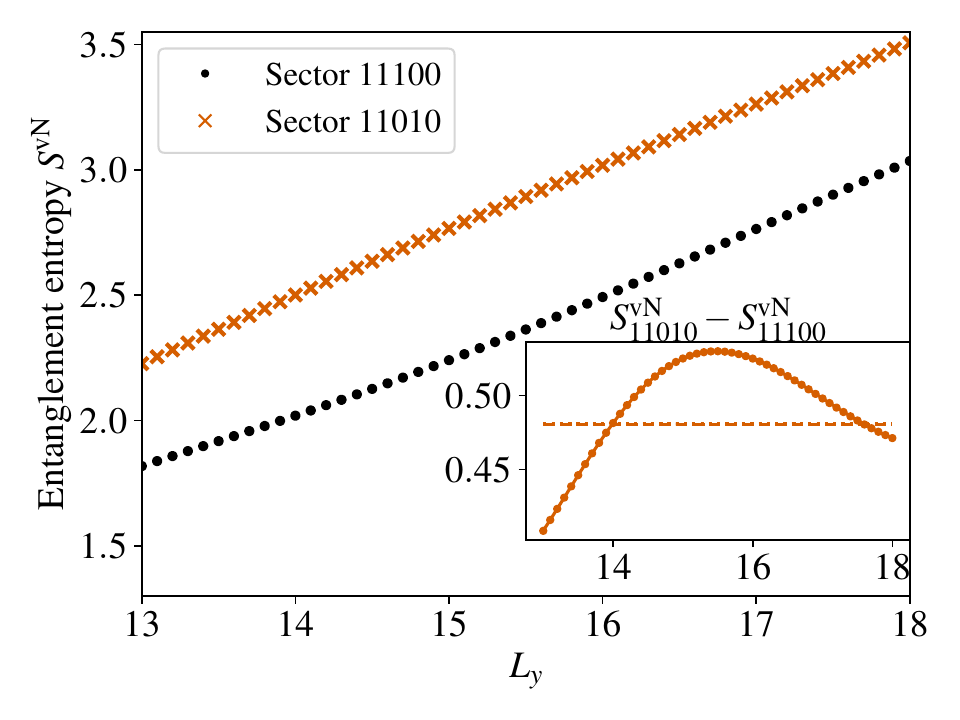}
\includegraphics[width=\linewidth]{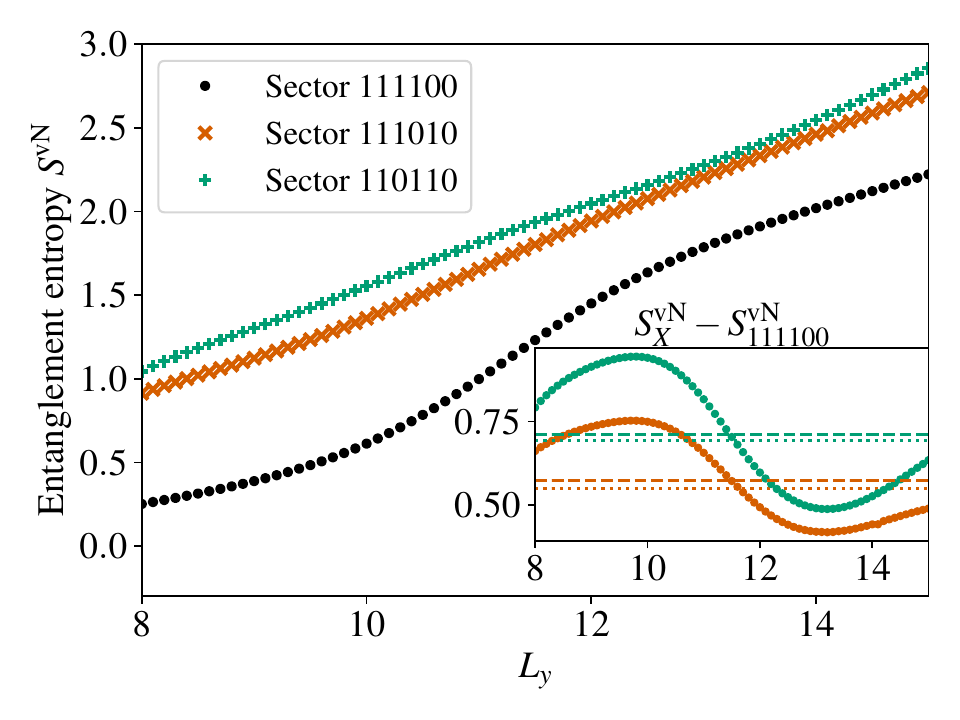}
\end{center}
\caption{Scaling of the entanglement entropy for the Moore-Read (top), RR35 (middle) and RR46 (bottom) states depending on the root configuration. Note that we represent the entanglement entropy averaged over all cuts in the unit-cell. In inset, we represent the difference in entropy between states with non-trivial boundary configurations and the reference trivial state. The large dots are the numerical data, the full line a fit using Eq.~\ref{eq:Fit}, the dashed line the extracted topological entanglement entropy and the dotted line the theoretical value. In all cases, we are able to extract the thermodynamic value with good precision. The small discontinuity in the inset for the Moore-Read Pfaffian is due to the iMPO truncation. It serves as a clear estimator on the precision of the obtained entanglement entropy. }
\label{fig:TEE}
\end{figure}

The groundstates of gapped, local Hamiltonians follow the area law, i.e., their entanglement grows with the boundary of $\mathcal{A}$.
In this paper, we consider states on an infinite cylinder, and $\mathcal{A}$ includes all orbitals to the left of a chosen cut.
Note that it is not a cut in real space, but in orbital space.
The entanglement entropy scales as
\begin{equation}
S^\mathrm{vN}(\mathcal{A}) = \alpha L_y - \gamma + o(1).
\end{equation}
$\gamma$ is the topological entanglement entropy\citep{Kitaev2006, Levin2006}, which partially characterizes long-range topological order.
On a cylinder, $\gamma = \log D / d_a$ where $D$ is the total quantum dimension of the theory ($D^2 = \sum\limits_a d_a^2 $) and $d_a$ is the quantum dimension of the edge excitation.
The quantum dimensions of the $\mathbb{Z}_k$ parafermions that appear in the Read-Rezayi series are
\begin{equation}
d_{m, k} = \frac{\sin \left[(m+1) \frac{\pi}{k+2} \right]}{\sin \frac{\pi}{k+2}}, \text{ with } m \in [0, k-1].
\end{equation}
For the Moore-Read state and the Majorana excitation $\gamma$, the entropies verify in the thermodynamic limit $S^\mathrm{vN}_{\gamma \times \gamma} - S^\mathrm{vN}_{1 \times 1} = \log \sqrt{2}$.
For the RR35 state and the $\mathbb{Z}_3$ excitation $\tau$, $S^\mathrm{vN}_{\tau \times \overline{\tau}} - S^\mathrm{vN}_{1 \times 1} = \log \phi$ where $\phi$ is the golden ratio\citep{Estienne2015}.
Finally, for the RR46 state and the $\mathbb{Z}_4$ excitation $\tau_4$, $S^\mathrm{vN}_{\tau_4 \times \overline{\tau_4}} - S^\mathrm{vN}_{1 \times 1} = \log \sqrt{3}$ and $S^\mathrm{vN}_{\tau_4^2 \times \overline{\tau_4}^2} - S^\mathrm{vN}_{1 \times 1} = \log 2$.
We verified that we recovered the expected values of the topological entanglement entropy for the different root configurations and the different Read-Rezayi states as shown in Fig.~\ref{fig:TEE}.
Due to the finite width of the cylinders, oscillations around the thermodynamic value are significant.
Correct estimates of the topological entropy can nonetheless be extracted using a fitting formula similar to Eq.~\eqref{eq:Fit}.
The residual errors are a combination of higher-order finite-size effects not taken into account in our fitting formula and of the truncation of the MPOs.

\end{document}